# Study of collective effects in the FCC-ee collider

**M Zobov[1], E Belli[2,3], G Castorina[3], M Migliorati[3], S Persichelli[4], G Rumolo[2] and B Spataro[1]**

[1]LNF-INFN, Via Enrico Fermi 40, 00044 Frascati (Rome), Italy
[2]CERN, Geneva, Switzerland
[3]University of Rome "La Sapienza" and INFN Sez. Roma1, Rome, Italy
[4]LBNL, Berkeley, California, USA

E-mail: mikhail.zobov@lnf.infn.it

**Abstract**. The Future Circular Collider (FCC) study aims at designing different options of a post-LHC collider. The high luminosity electron-positron collider FCC-ee based on the crab waist concept is considered as an intermediate step on the way towards FCC-hh, a 100 TeV hadron collider using the same tunnel of about 100 km. Due to a high intensity of circulating beams the impact of collective effects on FCC-ee performance has to be carefully analyzed. In this paper we evaluate beam coupling impedance of the FCC-ee vacuum chamber, estimate thresholds and rise times of eventual single- and multibunch beam instabilities and discuss possible measures to mitigate them.

## 1. Introduction

The first electron-positron Higgs factory was proposed already several months before the official announcement of the Higgs boson discovery [1]. It was a high luminosity collider in the LHC tunnel, called "LEP3", inspired by a low emittance B-factory design [2] and lepton-proton collider LHeC design studies [3]. The proposal was aimed at reaching the luminosity of the order of $10^{34}$ cm$^{-2}$s$^{-1}$ at the energy of 120 GeV per each beam for precise measurements of the Higgs boson mass, cross-section and its decay modes. The proposal has evolved, first passing to DLEP, the e+e- collider having the double LHC circumference, and then to TLEP ("Triple LEP") hosted in a 80 to 100 km tunnel [4]. The use of such a large tunnel has opened the possibility to accommodate also a hadron machine with an unprecedented energy of 100 TeV in the center-of-mass. The overall project is being implemented at CERN under the "Future Circular Collider" (FCC) design study [5] with FCC-ee being the electron-positron collider (former TLEP) considered as a potential first step towards the 100 TeV hadron machine called FCC-hh. The electron-proton collider option, FCC-eh, is also considered in the framework of these studies. FCC-ee is foreseen to operate at different center-of-mass energies ranging from 91 GeV to 365 GeV to study the properties of the Z resonance, the W and top pair thresholds and the Higgs boson with very high precision (see Table 1). For example, the design luminosity of 2.3x10$^{36}$ cm$^{-2}$s$^{-1}$ at Z resonance is almost 5 orders of magnitude higher than the maximum luminosity ever achieved at LEP at the same energy (see Table 2 in [6]). There are several ingredients that contribute in reaching such high luminosities in FCC-ee. These are some of them:

- Two separate rings allow to collide many bunches without their parasitic interactions.
- The longer perimeter allows to store higher beam intensities with the same synchrotron radiation (SR) loss.

- The crab waist collision scheme, proposed [7,8] and successfully tested at Frascati [9], allows to reduce the betatron functions at the interaction point, to collide beams with much lower emittances and to alleviate the nonlinear beam-beam interaction by suppressing beam-beam resonances [10].

Table 1. Relevant FCC-ee baseline parameters

| Parameter | Z | WW | H(ZH) | ttbar | |
|---|---|---|---|---|---|
| Circumference $C$ [km] | 97.75 | 97.75 | 97.75 | 97.75 | 97.75 |
| Energy $E$ [GeV] | 45.6 | 80 | 120 | 175 | 182.5 |
| Number of bunches per beam | 16640 | 2000 | 328 | 59 | 48 |
| Bunch population $N_p$ [$10^{11}$] | 1.7 | 1.5 | 1.8 | 2.2 | 2.3 |
| Beam current $I$ [mA] | 1390 | 147 | 29 | 6.4 | 5.4 |
| SR energy loss per turn [GeV] | 0.036 | 0.34 | 1.72 | 7.8 | 9.21 |
| Bunch length with SR/BS, $\sigma_z$ [mm] | 3.5/12.1 | 3.0/6.0 | 3.15/5.3 | 2.75/3.82 | 1.97/2.54 |
| Bunch energy spread, SR/BS [%] | 0.038/0.132 | 0.066/0.131 | 0.099/0.165 | 0.144/0.196 | 0.150/0.192 |
| Longitudinal damping time [turns] | 1281 | 235 | 70 | 23.1 | 20 |
| Horizontal emittance $\varepsilon_x$ [nm] | 0.27 | 0.84 | 0.63 | 1.34 | 1.46 |
| Vertical emittance $\varepsilon_y$ [pm] | 1.0 | 1.7 | 1.3 | 2.7 | 2.9 |
| Luminosity per IP [$10^{34}$ cm$^{-2}$s$^{-1}$] | 230 | 28 | 8.5 | 1.8 | 1.55 |

Analyzing the parameters shown in Table 1 we can see that the beam emittances of FCC-ee are very small, comparable to those of the modern synchrotron light sources, while the beam stored currents are close to the best current values achieved in the last generation particle factories (see Table 2 in [11] for comparison). Therefore, a careful study of collective effects is required in order to preserve the quality of the intense beams, to suppress eventual beam instabilities and to avoid excessive RF power losses leading to a damage of vacuum chamber components and accelerator hardware.

In this paper we present a preliminary study of the collective effects in FCC-ee due to parasitic electromagnetic interaction of the circulating beams with the surrounding vacuum chamber expressed in terms of wake fields and beam coupling impedances [12]. We will consider only the Z resonance option since it is more vulnerable to collective effects and instabilities because of the lower beam energy, longer damping times, higher beam intensities and highest number of bunches.

## 2. Beam coupling impedance

As it has been shown in [13], for the 100 km long collider, the finite conductivity of the beam vacuum chamber represents a major source of the beam coupling impedance, and wake fields strongly affect the beam dynamics and, respectively, collider design solutions and parameters choice. This fact has been taking into account while choosing the vacuum chamber shape and its dimensions. In particular, it has been proposed to use the vacuum chamber with a round cross-section in order to avoid the betatron tune shift with the beam current in multibunch operations due to quadrupolar resistive wall (RW) wake fields [14, 15]. The beam pipe radius of 35 mm has been chosen as a compromise between the beam impedance and the power required for magnet power supplies. Actually the shape of the pipe is not totally round but additional antechambers (winglets) are foreseen for pumping purposes and installation of SR absorbers, similarly to SuperKEKB design [16]. The twin dipole and quadrupole magnets of FCC-ee are being designed incorporating this vacuum chamber cross-section [17]. The constant vacuum chamber cross-section helps to avoid the use of multiple tapered transitions between different vacuum chamber sections, thus further improving the machine impedance.

The FCC-ee vacuum chambers have to be coated in order to mitigate the electron cloud effects (beam induced multipacting) in the positron ring and/or to improve the vacuum pumping in both rings. Thin layers of NEG, TiN and AC have been considered for these purposes [18]. It has been demonstrated [13] that under certain assumptions, that are valid for the FCC-ee parameters, the longitudinal and transverse impedances of a two-layer beam pipe are given by the sum of two terms, the first term representing the well-known impedance of a single layer beam pipe and the second one describing an

inductive perturbation proportional to the thickness of the coating:

$$\frac{Z_L(\omega)}{C} \approx \frac{Z_0 \omega}{4\pi c b}\left\{[sgn(\omega) - i]\delta_2 - 2i\Delta\left(1 - \frac{\sigma_1}{\sigma_2}\right)\right\} \quad 1$$

$$\frac{Z_T(\omega)}{C} \approx \frac{Z_0}{2\pi b^3}\left\{[1 - i\,\text{sgn}(\omega)]\delta_2 - 2i\Delta\,\text{sgn}(\omega)\left(1 - \frac{\sigma_1}{\sigma_2}\right)\right\} \quad 2$$

Where $Z_0$ is the vacuum impedance, $c$ the speed of light, $b$ the pipe radius, $\delta_1$, $\sigma_1$ and $\delta_2$, $\sigma_2$ the skin depths and conductivities of the coating and the beam pipe (substrate), respectively.

As it can be seen, for the above assumptions, the real part of the impedance does not depend on the coating thickness and conductivity. The performed numerical studies [13] have confirmed that the resulting RF power losses due to the RW impedance remain almost unchanged for the coating conductivity and thickness varying in a very wide range. In turn, since the perturbation of the imaginary part due to the coating is proportional to its thickness and to the term $\left(1 - \frac{\sigma_1}{\sigma_2}\right)$, if the coating conductivity is much smaller than the beam pipe material conductivity the impedance depends only on the thickness of the coating layer. On the other hand, if the two conductivities are comparable, this term reduces the coating perturbation. It has been shown that the coating thickness plays a crucial role affecting both single and multibunch beam dynamics [13]. In particular, in order to keep longitudinal microwave instability under control the coating thickness should be as small as 50-100 nm [19]. For this reason, a campaign of dedicated measurements has been launched to study the properties of NEG thin films with thicknesses below 250 nm [19, 20], such as secondary emission yield and activation performance.

Table 2. RF power losses due to different vacuum chamber components

| **Component** | **Number** | **$k_{loss}$ [V/pC]** | **$P_{loss}$ [MW]** |
|---|---|---|---|
| Resistive wall | 97.75 km | 210 | 7.95 |
| Collimators | 20 | 18.69 | 0.7 |
| RF cavities | 52 | 17.14 | 0.65 |
| RF double tapers | 13 | 24.71 | 0.93 |
| Beam position monitors | 4000 | 40.11 | 1.50 |
| Bellows | 8000 | 49.01 | 1.85 |
| **Total** | | **359.6** | **13.6** |

In addition to the RW impedance there are many other impedance sources in the machine. The design of the vacuum chamber components such as RF cavities, kickers, beam position monitors (BPMs), bellows, flanges etc. has not been completed yet. In order to evaluate their possible impedance contribution, the present strategy consists in adopting the best design solutions of the accelerator components used in the modern synchrotron light sources and the particle factories. For the present impedance model we consider that at Z running the RF system consists of 52 single cell cavities operating at 400 MHz [21] and arranged in groups of four cavities connected to the beam pipe by 26 0.5 m long tapers. In order to eliminate the beam halo and to suppress the background, collimators based on PEP-II and SuperKEKB design [22, 23] are planned to be installed in the machine, for a total number of 20 (10 for each plane). The impedance contribution of the 10000 absorbers to cope with the SR has been minimized by placing them inside the two rectangular antechambers on both sides of the beam pipe. This model also includes 4000 BPMs [24] and 8000 comb-type bellows with RF shielding [25] to be allocated before and after each BPM.

The coupling impedances and wake fields for these vacuum chamber elements have been evaluated numerically. Figure 1 shows the longitudinal wake potentials of each component for the nominal bunch length of 3.5 mm. Table 2 summarizes the corresponding loss factors. As it can be seen, the resistive walls with 100 nm coating provide the dominating contribution in both the total wake potential and respective power losses that are not negligible compared with the 50 MW power lost due to SR.

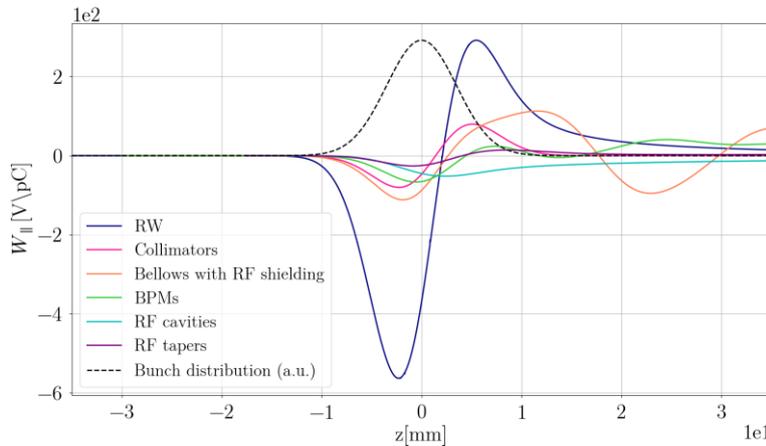
Figure 1 Longitudinal wake potential for different vacuum chamber components calculated for 3.5 mm Gaussian bunch

### 3. Collective effects issues

The electromagnetic beam interaction with a surrounding vacuum chamber, described in terms of wake fields and impedances, affects longitudinal and transverse beam dynamics. It can result in both single and multi-bunch instabilities and overheating of vacuum chamber components. The longitudinal wake potential (see Fig.1) modifies the RF voltage leading to a bunch length increase and a bunch shape distortion. In collisions with a large Piwinski angle, as is the case of FCC-ee, the collider geometric luminosity and beam-beam interaction depend on the bunch length, thus influencing the overall collider performance. For longer bunches the beam lifetime increases and the RF power losses are reduced. However, at a certain bunch intensity, microwave instability can take place. Typically the microwave instability does not produce a bunch loss, but the consequent energy spread growth and possible bunch internal oscillations above the instability threshold cannot be counteracted by a feedback system. In addition, the longitudinal wake fields result in the synchrotron tune reduction and a large incoherent synchrotron tune spread. Both these effects influence beam-beam performance shifting the collider working point and affecting the coherent and incoherent beam-beam resonances. The left plots in Fig.2 show the bunch length in FCC-ee as a function of the bunch population, while the right plots indicate the respective energy spread. The blue curves correspond to the bunch length and energy spread variations for non-colliding bunches. At the nominal intensity the bunch lengthens till about 7 mm, while the microwave instability threshold is about a factor of 1.5 higher than the nominal bunch population. So there is only a small margin left for eventual impedance increase. In collision the "beamstrahlung" effect [26] results in an additional energy spread increase leading to the strong bunch elongation and the microwave instability threshold increase beyond the considered bunch intensities (brown curves).

Differently from the longitudinal microwave instability, the transverse mode coupling instability (TMCI) is destructive for intense bunches. It takes place when coherent frequencies of different modes of transverse internal bunch oscillations merge. The TMCI threshold has been evaluated with the analytical Vlasov solver DELPHI [27] by considering the dominating RW impedance and by taking into account the bunch lengthening due to the longitudinal wake fields shown in Fig.2. It has been found that in the transverse case the TMCI instability threshold is affected to a lesser extent by the coating thickness due to the bunch lengthening effect. For comparison, Fig. 3 shows the real part of the frequency shift of the first coherent oscillation modes as a function of the bunch population for 50 nm (left) and 1 µm coatings (right), respectively. The dashed line represents the nominal bunch intensity. The TMCI threshold (merging lines) is about a factor 2.5 higher than the nominal intensity.

Analyzing the multi-bunch beam dynamics it has been found that the coupled bunch instability due to the transverse RW long range wake fields is another critical issue for the collider [13]. The growth rate of the fastest coupled bunch mode is estimated to be 435 $s^{-1}$, corresponding to about 7 revolution turns.

It is worth noting here that the transverse radiation damping time is 2550 turns, i.e. it is much longer than required for the instability suppression. So a robust feedback system is necessary to mitigate the fast instability. A dedicated study is underway to develop such a challenging feedback system [28].

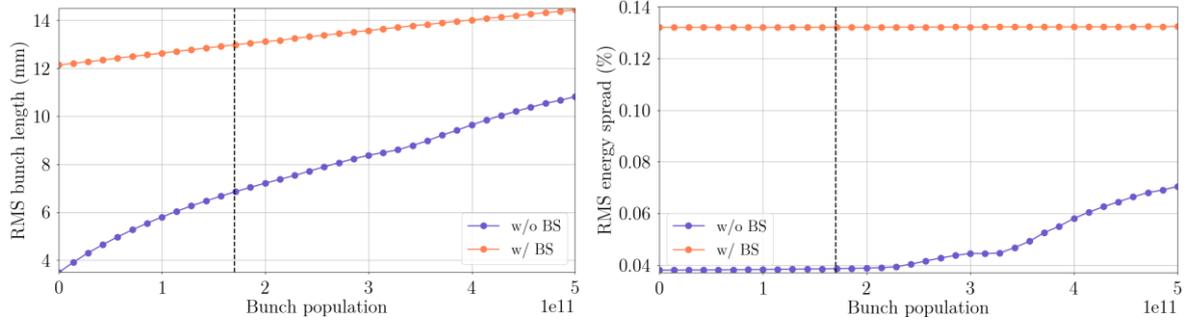

Figure 2 Bunch lengthening (on the left) and energy spread (on the right) in FCC-ee.

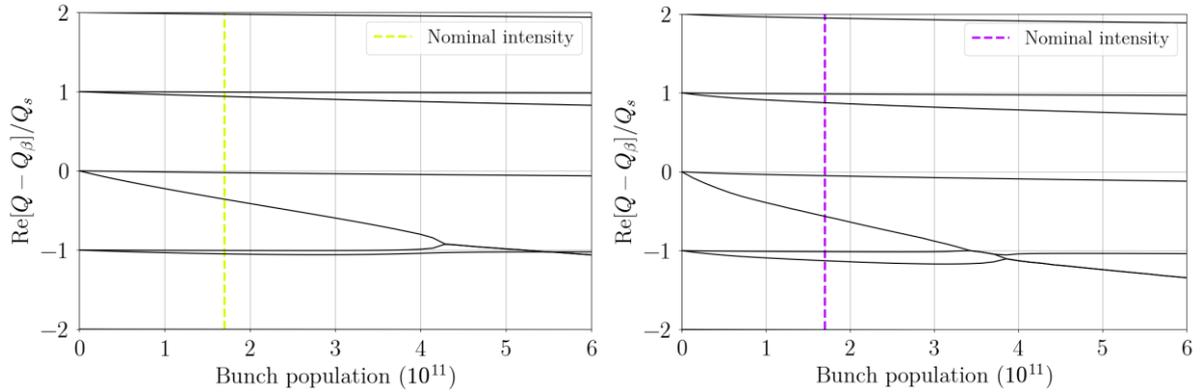

Figure 3 Real part of the frequency shift of the first transverse coherent oscillation modes for 50 nm (left) and 1 μm (right) coating thickness

The longitudinal radiation damping alone also cannot suppress the longitudinal coupled bunch instabilities due to the beam interaction with parasitic higher order modes (HOM) trapped in the vacuum chamber components. In order to cope with the instabilities special HOM damping techniques are to be applied to reduce the shunt impedances of the HOM to a harmless level, as discussed in [28]. In addition, also in this case a longitudinal feedback system has to be developed as a further safety knob.

## 4. Conclusions
FCC-ee beam impedance plays an important role for both the parameters choice and design solutions for the 100 km collider. In particular it has been shown that the impedance has an impact on the choice of the vacuum chamber shape, properties of the beam pipe coating, requirements of the feedback systems, and defines the microwave and TMCI instability thresholds, bunch lengthening, energy spread and RF energy loss. Further work is in progress to design low impedance vacuum chamber components and to investigate other collective effects such as the beam induced electron cloud, ion instabilities etc.

**Acknowledgements**
This work was in part supported by the European Commission under the HORIZON2020 Integrating Activity project ARIES, grant agreement 730871, and in part by the MICA project of INFN.